\documentclass[fleqn,usenatbib]{mnras}

\usepackage{newtxtext,newtxmath}

\usepackage[T1]{fontenc}

\DeclareRobustCommand{\VAN}[3]{#2}
\let\VANthebibliography\thebibliography
\def\thebibliography{\DeclareRobustCommand{\VAN}[3]{##3}\VANthebibliography}

\usepackage{graphicx}	
\usepackage{amsmath}	
\usepackage{amssymb}	

\usepackage{cleveref}
\usepackage{soul}
\usepackage{xcolor}

\title[$H_0$ measurement from FRBs]
{A new measurement of the Hubble constant using Fast Radio Bursts}
\author[Hagstotz, Reischke and Lilow]
{
Steffen Hagstotz\thanks{E-mail: \href{mailto:steffen.hagstotz@fysik.su.se}{steffen.hagstotz@fysik.su.se}}$^1$, Robert Reischke\thanks{E-mail:  \href{mailto:reischke@astro.ruhr-uni-bochum.de}{reischke@astro.ruhr-uni-bochum.de}}$^{2}$ and Robert Lilow\thanks{E-mail: \href{mailto:rlilow@campus.technion.ac.il}{rlilow@campus.technion.ac.il}}$^{3}$
\\
$^1$ The Oskar Klein Centre for Cosmoparticle Physics,  Department of Physics, Stockholm University, Roslagstullsbacken 21A, SE-106 91 Stockholm, Sweden
\\
$^2$ Ruhr University Bochum, Faculty of Physics and Astronomy, Astronomical Institute (AIRUB),\\ \hspace{0.15cm} German Centre for Cosmological Lensing, 44780 Bochum, Germany \\
$^3$ Department of Physics, Technion, Haifa 3200003, Israel
\\
} 

\date{Accepted XXX. Received YYY; in original form ZZZ}

\pubyear{2021}

\begin{document}
\label{firstpage}
\pagerange{\pageref{firstpage}--\pageref{lastpage}}
\maketitle

\begin{abstract}
Fast radio bursts (FRBs) are very short and bright transients visible over extragalactic distances. The radio pulse undergoes dispersion caused by free electrons along the line of sight, most of which are associated with the large-scale structure (LSS). The total dispersion measure therefore increases with the line of sight and provides a distance estimate to the source.
We present the first measurement of the Hubble constant using the dispersion measure -- redshift relation of FRBs with identified host counterpart and corresponding redshift information. A sample of nine currently available FRBs yields a constraint of $H_0 = 62.3 \pm 9.1 \,\rm{km} \,\rm{s}^{-1}\,\rm{Mpc}^{-1}$, accounting for uncertainty stemming from the LSS, host halo and Milky Way contributions to the observed dispersion measure.
The main current limitation is statistical, and we estimate that a few hundred events with corresponding redshifts are sufficient for a per cent measurement of $H_0$. This is a number well within reach of ongoing FRB searches. We perform a forecast using a realistic mock sample to demonstrate that a high-precision measurement of the expansion rate is possible without relying on other cosmological probes. FRBs can therefore arbitrate the current tension between early and late time measurements of $H_0$ in the near future.\end{abstract}




\section{Introduction}

Fast radio bursts (FRBs) are very short (ms) transients observed in frequencies from $\sim 100$ MHz up to a few GHz.
The mechanism causing the burst is still unknown, but at least some of them are associated with magnetars \citep{Bochenek:2020zxn}.
A subset of bursts have been found to repeat, either with a fixed period \citep{Amiri:2020gno} or in cyclical phases of irregular activity \citep{Rajwade:2020uat}. It is therefore possible that the observed FRB events actually fall into a mix of different populations or progenitor mechanisms.
The observed radio bursts are bright enough to be visible over extragalactic distances, which opens up exciting possibilities to use FRBs for studying cosmological scales.

Interest in FRBs has surged in the past years, and after the first accidental discovery in archival data in 2007 \citep{Lorimer:2007qn}, more than hundred events have been observed by radio telescopes all around the world \citep[see][for a compilation]{Petroff:2016tcr}.
Many dedicated search programs are now ongoing or about to start. These will increase the number of known events by many orders of magnitude. Recently, the Canadian Hydrogen Intensity Mapping Experiment \citep[CHIME,][]{CHIME:2014}, the Hydrogen Intensity and Real-time Analysis eXperiment \citep[HIRAX,][]{HIRAX:2016} and the Australian Square Kilometre Array Pathfinder \citep[ASKAP,][]{ASKAP:2007yh} started taking data, and all of them are expected to detect tens of events every night. Eventually, large-scale experiments such as the Square Kilometre Array \citep[SKA,][]{SKA:2018lat} are going to discover up to thousands of FRBs per night.

The radio pulse from the burst undergoes dispersion while travelling through the ionized intergalactic medium, and the total inferred dispersion measure (DM) is a powerful probe of the column density of ionised electrons along the line of sight \citep{thornton_population_2013, petroff_real-time_2015, connor_non-cosmological_2016, champion_five_2016,chatterjee_direct_2017}. The observed DM exceeds the expected values from the Milky Way halo by far, which together with the isotropic distribution of events on the sky strongly suggests an extragalactic origin of most observed FRBs. Determining the redshift by identifying the host galaxy of the burst has been challenging so far and only nine events have been localised to sufficient accuracy. However, the ongoing surveys use interferometry to achieve very large effective baselines, and thus the resulting enhanced angular resolution will improve the situation considerably. Together with dedicated follow-up programs to measure the redshift of repeating FRBs such as the Deep Synoptic Array \citep[DSA,][]{DSA10:2019MNRAS}, we can expect redshift information for a substantial subset of the newly discovered FRBs.

The large future samples of FRBs also make it possible to use DM correlations to probe the perturbations in the electron distribution on very large scales \citep{masui_dispersion_2015,shirasaki_large-scale_2017,rafiei-ravandi_characterizing_2020,bhattacharya_fast_2020,takahashi_statistical_2021}. This provides a unique way to, for example, constrain primordial non-Gaussianity \citep{Reischke:2020cgd} or deviations from General Relativity \citep{Reischke:2021euf}. For the current small numbers of FRBs, however, the correlation signal is dominated by shot noise. We will therefore focus on the total DM of the bursts directly.

Since the DM is a noisy distance estimate, FRBs with known host can be used to compile a DM -- redshift relation. This turns FRBs into standardisable radio transmitters, which can be used to probe the baryon content of the Universe at late times \citep{Walters:2019cie, Macquart:2020lln}, the ionisation history of the IGM \citep{Jaroszynski:2018vgh, bhattacharya_fast_2020, Pagano:2021zla} or the current expansion rate \citep{Wu:2020jmx}.

In this paper, we follow the latter possibility and use the currently nine FRBs with known redshift to measure the Hubble constant $H_0$ from the $\mathrm{DM}(z)$ relation. Currently, the determination of $H_0$ by Cepheid-calibrated Supernovae Ia \citep{Riess:2019cxk} is in strong tension with measurements based on the observation of the imprint left by baryon acoustic oscillations in the CMB \citep{Aghanim:2018eyx} or in the large-scale structure \citep{Lemos:2018smw}. Despite a lot of effort, no convincing theoretical solution to the tension has been found so far \citep[see e.g.][for a summary]{Knox:2019rjx}. Methods that can determine $H_0$ without using either the BAO scale or the direct distance ladder can potentially shed light on the problem. Alternative determinations of $H_0$ based on Supernovae Ia calibrated using the tip of the red giant branch \citep{Freedman:2019jwv}, strongly lensed quasars \citep{Wong:2019kwg, Birrer:2020tax}, MASERs \citep{pesce_megamaser_2020} or standard sirens \citep{Abbott:2017xzu} have so far marginally favoured one or the other result, but a clear picture is yet to emerge.

In this paper, we present the first measurement of the Hubble constant from the DM -- redshift relation of localised FRBs. In \cref{sec:DM} we introduce the necessary theoretical background for the calibration of the relation, followed by an overview over our dataset and the results in \cref{sec:results}. We explore the exciting possibilities of $H_0$ measurements based on the FRB samples available in the near future in \cref{sec:forecast}, before concluding with a discussion of the results in \cref{sec:conclusion}.


\section{Dispersion measure of FRBs}
\label{sec:DM}

The radio pulse from the FRB undergoes dispersion when travelling through the ionised intergalactic medium. The consequence is a characteristic time delay between the arrival times of different pulse frequencies $\nu$,
\begin{equation}
\label{eq:time_delay}
    \Delta t \propto \mathrm{DM}(\hat{\boldsymbol
    {x}}, z) \, \nu^{-2} \, ,
\end{equation}
which defines the inferred dispersion measure DM of a FRB at the sky position $\hat{\boldsymbol{x}}$.
This is caused by the free electrons along the line of sight. These electrons are either associated with the host halo, with the Milky Way, or with the large-scale structure (LSS). We therefore split the total contribution up to write
\begin{equation}
    \mathrm{DM}_\mathrm{tot}(\hat{\boldsymbol
    {x}}, z) = \mathrm{DM}_\mathrm{host}(z) + \mathrm{DM}_\mathrm{MW}(\hat{\boldsymbol
    {x}}) + \mathrm{DM}_\mathrm{LSS}(z) \; ,
\end{equation}
and we will discuss the various contributions and their respective redshift scaling one by one.

\subsection{Dispersion measure from diffuse electrons}
\label{subsec:DM_LSS} 

The average contribution from diffuse electrons in the LSS to the DM can be written as
\begin{equation}
\label{eq:DM_LSS}
    \mathrm{DM}_\mathrm{LSS}(z) = \int_0^z \! n_\mathrm{e}(z') \, f_\mathrm{IGM}(z') \, \frac{1+z'}{H(z')} \, \mathrm d z' \; ,
\end{equation}
where $n_\mathrm{e}$ is the cosmic free electron density, and the scaling with the expansion rate $H(z)$ makes the expression sensitive to the Hubble constant $H_0$. Since we are accounting for contributions from nonlinear objects separately, the overall DM is scaled by the fraction $f_\mathrm{IGM}(z)$ of electrons in the IGM that are not bound in structures. Since at redshifts $z<3$ almost all baryons are ionised, the cosmic electron density can be expressed as a function of the overall baryon abundance,
\begin{align}
\label{eq:n_e}
    n_\mathrm{e}(z) &= \int \! \chi_\mathrm{e} \frac{\rho_\mathrm{b}(\boldsymbol{x}, z)}{m_\mathrm{p}} \, \mathrm{d} \Omega = \int \! \chi_\mathrm{e} \frac{\bar \rho_\mathrm{b}}{m_\mathrm{p}} \big( 1 + \delta_\mathrm{e} (\boldsymbol{x}, z) \big) \, \mathrm{d} \Omega \\
    &\approx \chi_\mathrm{e} \frac{\bar \rho_\mathrm{b}}{m_\mathrm{p}} \; ,
\end{align}
with the baryon density $\rho_\mathrm{b}$, the proton mass $m_\mathrm{p}$ and the electron fraction
\begin{align}
\label{eq:chi_e}
    \chi_\mathrm{e} &= Y_\mathrm{H} + \frac{1}{2} Y_\mathrm{He} \\
    & \approx 1 - \frac{1}{2} {Y}_\mathrm{He} \, ,
\end{align}
calculated from the primordial hydrogen and helium abundances $Y_\mathrm{H}$ and $Y_\mathrm{He}$. In this paper, we assume $Y_\mathrm{H} \approx 1 - Y_\mathrm{He}$ and $Y_\mathrm{He} = 0.24$, found to high precision both by CMB measurements \citep{Aghanim:2018eyx} and by spectroscopic observations of metal-poor gas clouds \citep{Aver:2015iza}.

The baryon number density in \cref{eq:n_e} can be expanded around its mean background value $\bar \rho_\mathrm{b} / m_\mathrm{p}$ with the relative electron overdensity $\delta_\mathrm{e}$. The DM can therefore be used to probe the matter perturbations on large scales, but this requires a larger sample of FRBs than currently available. We will therefore stick to the average expression for this paper. However, large future FRB samples will open up unique opportunities to probe the electron density on large scales with DM correlations.

The fraction of electrons in the IGM in \cref{eq:DM_LSS} can be calculated by subtracting the fraction bound in stars, compact objects and the dense interstellar medium (ISM)
\begin{equation}
\label{eq:f_IGM}
    f_\mathrm{IGM}(z) = 1 - f_\star(z) - f_\mathrm{ISM}(z) \, .
\end{equation}
We compute\footnote{The code for the calculations is publicly available at \href{https://github.com/FRBs/FRB}{https://github.com/FRBs/FRB}, provided by \cite{Macquart:2020lln}.} $f_\star$ and $f_\mathrm{ISM}$ using the estimates of star formation rate and ISM mass fraction from \cite{Fukugita:2004ee, Madau:2014bja}. Since there is little star formation at low redshifts, the overall fraction of electrons in the IGM does not evolve significantly over the redshift range covered by the FRB sample, and we keep $f_\mathrm{IGM} = 0.84$ constant for the purposes of this analysis.

Putting everything together, we can rewrite the mean DM -- redshift relation in \cref{eq:DM_LSS} as
\begin{equation}
\label{eq:DM_LSS_v2}
    \mathrm{DM}_\mathrm{LSS}(z) = \frac{3 \Omega_\mathrm{b} H_0}{8 \pi G m_\mathrm{p}} \chi_\mathrm{e} \, f_\mathrm{IGM} \int_0^z \, \frac{1+z'}{E(z')} \mathrm d z' \; ,
\end{equation}
with the dimensionless baryon density parameter $\Omega_\mathrm{b}$ and the dimensionless expansion function $E(z) = H(z)/H_0$. This makes it clear that the FRB DM -- redshift relation at the background level is mostly sensitive to the product $\Omega_\mathrm{b} H_0$, with a little sensitivity to $\Omega_\mathrm{m}$ via the integral over $E(z)$.
Since most of the baryons in the late Universe are located in low-density, faint filaments, they are very difficult to find with other methods. For a while, these missing baryons at late times were puzzling \citep{2012ApJ...759...23S} until an analysis using the average DM of FRBs located them \citep{Macquart:2020lln}.

FRBs can also probe the baryon feedback properties via $f_\mathrm{IGM}$ or the electron ionisation fraction at high redshifts given independent measurements of the other quantities. Note that at the background level, every parameter combination that leaves the prefactor of the integral in \cref{eq:DM_LSS_v2} unchanged is perfectly degenerate.

The distribution of electrons in the IGM is inhomogeneous, and measurements in hydrodynamical simulations \citep{Jaroszynski:2018vgh} show that there is a stochastic contribution to the LSS dispersion measure. This component is well described by a Gaussian around the mean given by \cref{eq:DM_LSS_v2}, and we interpolate the standard deviation linearly from the values found in simulations, using $\sigma_\mathrm{LSS}(z=0) \approx 40 \, \mathrm{pc} \, \mathrm{cm}^{-3}$ and $\sigma_\mathrm{LSS}(z=1) \approx 180 \, \mathrm{pc} \, \mathrm{cm}^{-3}$. The scatter was measured in the Illustris simulations \citep{Vogelsberger:2014kha}, and we expect it to scale proportionally to $\Omega_\mathrm{b} h$, in the same way as the mean relation, when the cosmological parameters are varied.

\subsection{Host and Milky Way contribution}
\label{subsec:DM_structures}

The discussion so far only assumes the DM at the background level and cannot account for highly non-linear structures. For typical FRBs, there are two of those objects along the line of sight: the host halo and the Milky Way.

The Milky Way DM can be predicted and removed with the help of models of the galactic electron distribution. We follow \cite{Petroff:2016tcr} and use the NE2001 model \citep{Cordes:2002wz} to subtract the Milky Way contribution for each FRB position in the sky. The predicted DM values are in agreement with pulsar measurements up to $\sigma_\mathrm{MW} \approx 30\: \mathrm{pc}\,\mathrm{cm}^{-3}$ \citep{Manchester:2004bp}, which we take as a measure for the uncertainty of the model.

The host galaxy properties are more uncertain, and we assume a stochastic contribution
\begin{equation}
    p(\mathrm{DM}_\mathrm{host}) = \mathcal{N} \big( \langle \mathrm{DM}_\mathrm{host} \rangle, \sigma_\mathrm{host}^2 \big) \; ,
\end{equation}
where $\mathcal{N}$ denotes the normal distribution with given mean and variance. There is some uncertainty considering the range of host DM values, but for the purposes of this analysis we assume halos somewhat similar to the Milky Way with $\langle \mathrm{DM}_\mathrm{host} \rangle = 100 (1+z_\mathrm{host})^{-1} \, \mathrm{pc}\,\mathrm{cm}^{-3}$, and allow for a large scatter $\sigma_\mathrm{host} = 50 (1+z_\mathrm{host})^{-1} \, \mathrm{pc}\,\mathrm{cm}^{-3}$. Note that since \cref{eq:time_delay} applies to the rest frame, both mean and standard deviation of the observed host DM decay with increasing host redshift.

While there could be other halos along the line of sight, this is very rare and explicitly excluded by the optical host identification.

\section{Results}
\label{sec:results}

\subsection{Data}
\label{subsec:data}

Of the 118 verified FRBs publicly available, nine have a localised host galaxy and a corresponding redshift estimate. All of them are presented in \cref{tab:FRB_data}.  Six FRBs were detected and localised by ASKAP \citep{Prochaska_2019Sci, Bannister_2019Sci, Macquart:2020lln}.
FRB 180916 was the first localised repeating FRB \citep{Marcote:2020ljw} with a period of $\sim 16$ days \citep{Amiri:2020gno}. The source of the burst is located within a nearby galaxy at $z=0.0337$.

FRB 121102 was discovered by the Arecibo telescope \citep{Spitler_2014ApJ}, and later found to repeat with somewhat irregular periods of activity which form cycles of $\sim 157$ days \citep{Rajwade:2020uat}. These repeated emissions made it possible to identify the host later \citep{chatterjee_direct_2017, Tendulkar_2017ApJ} as a dwarf galaxy at $z=0.192$.

FRB 190523 was discovered by DSA-10 \citep{Ravi:2019alc} and is the most distant burst, with a host redshift of $z=0.66$, even though events have been recorded with substantially larger dispersion measures. This implies an FRB population beyond $z\sim 1$ detectable for the ongoing search programs.

Not all host identifications currently have the same degree of certainty. 
If the ASKAP FRB 190611 host is identified correctly, the FRB is displaced from the host galaxy by $\sim 10$ kpc, so the association is tentative \citep{Macquart:2020lln}. Both events FRB 180916 and FRB 12110 lie within the galactic disc, where the contamination from the Milky Way might be substantially higher than predicted by the NE2001 model. We therefore perform the analysis either with all available events, or limit ourselves to a gold sample where these three FRBs are excluded. The difference in results, however, is very minor.


\begin{table}
	\centering
	\caption{Overview of all FRBs with redshift localisation. The Milky Way DM is predicted for the sky position by the NE2001 model of the galactic electron distribution. The last digit in brackets of the measured DM indicates the uncertainty, which is always negligible compared to the systematic uncertainties. The FRBs marked with a star are excluded from the gold sample.}
	\label{tab:FRB_data}
	\begin{tabular}{cccc}
		\hline
		name & redshift $z$ & measured DM & Milky Way DM \\
		     &          & $[\mathrm{pc} \; \mathrm{cm}^{-3}]$ & $[\mathrm{pc} \; \mathrm{cm}^{-3}]$ \\
		\hline
 FRB 180916$^\star$          & 0.0337        &  348.8(1) & 199 \\
 FRB 190608$^\star$          & 0.1178        & 338.7(5)   & 37.2 \\
 FRB 121102\hphantom{$^\star$}          & 0.19273       &  558(3)  & 188  \\
 FRB 190102\hphantom{$^\star$}          & 0.291         &  363.6(3)  & 57.3 \\
 FRB 180924\hphantom{$^\star$}          & 0.3214        &  361.42(6) & 40.5 \\
 FRB 190611$^\star$          & 0.378         & 321.4(2)   & 57.8 \\
 FRB 181112\hphantom{$^\star$}          & 0.4755        &  589.27(3) & 102 \\
 FRB 190711\hphantom{$^\star$}          & 0.522         & 593.1(4)  & 56.4 \\
 FRB 190523\hphantom{$^\star$}          & 0.66          &  760.8(6)  & 37  \\
 \hline
	\end{tabular}
\end{table}

\subsection{Likelihood analysis}
\label{subsec:likelihood}

For the data analysis, we assume Gaussian individual likelihoods to observe a dispersion measure $\mathrm{DM}_i$ at a given redshift $z_i$,
\begin{equation}
\label{eq:likelihood_single}
    \mathcal{L}(\mathrm{DM}_i, z_i) = \frac{1}{\sqrt{2 \pi \sigma_i^2}} \exp \left( \frac{\bigl(\mathrm{DM}_i - \mathrm{DM}^\mathrm{theo}(z_i) \bigr)^2} {2 \sigma_i^2} \right) \,,
\end{equation}
with the theoretical prediction for the DM contributions as discussed in \cref{sec:DM},
\begin{equation}
    \mathrm{DM}^\mathrm{theo}(z_i) = \mathrm{DM}_\mathrm{LSS} (z_i) + \langle \mathrm{DM}_\mathrm{host} \rangle (z_i) \, .
\end{equation}
The measurement error on $\mathrm{DM}_i$ is negligible; thus the total variance follows from the individual uncertainties accounting for the scatter of the LSS contribution, the MW electron distribution model and the host galaxy,
\begin{equation}
\label{eq:scatter_tot}
    \sigma^2(z_i) =  \sigma_{\mathrm{MW}}^2 + \sigma_\mathrm{host}^2(z_i) + \sigma_\mathrm{LSS}^2(z_i) \, .
\end{equation}
The estimate for the individual scatter contributions is described in \cref{sec:DM}. Since all events are independent\footnote{The events FRB 190102, FRB 190611 and FRB 190711 are very close together in angular projection on the sky. The modelled Milky Way DMs and their error for these events are thus not entirely independent. However, this is a subdominant contribution to the scatter.}, the joint likelihood of the sample is then the product of the individual likelihoods,
\begin{equation}
\label{eq:likelihood_tot}
    \mathcal{L}_\mathrm{tot} = \prod_i \mathcal{L}_i \: ,
\end{equation}
and the product is performed either over all FRBs listed in \cref{tab:FRB_data}, or the gold sample alone.

For the analysis, we fix the physical densities $\Omega_\mathrm{m} h^2 = 0.143 $ and $\Omega_\mathrm{b} h^2 = 0.02237$ to the best-fit values found by the latest Planck CMB analysis \citep{Aghanim:2018eyx}. Their respective uncertainty is small compared to the statistical error resulting from the limited number of available FRBs. Since the DM -- redshift relation in \cref{eq:DM_LSS_v2} is proportional to the product $\Omega_\mathrm{b} h$, this makes our analysis sensitive to $h^{-1}$.

\begin{figure}
	\includegraphics[width=\columnwidth]{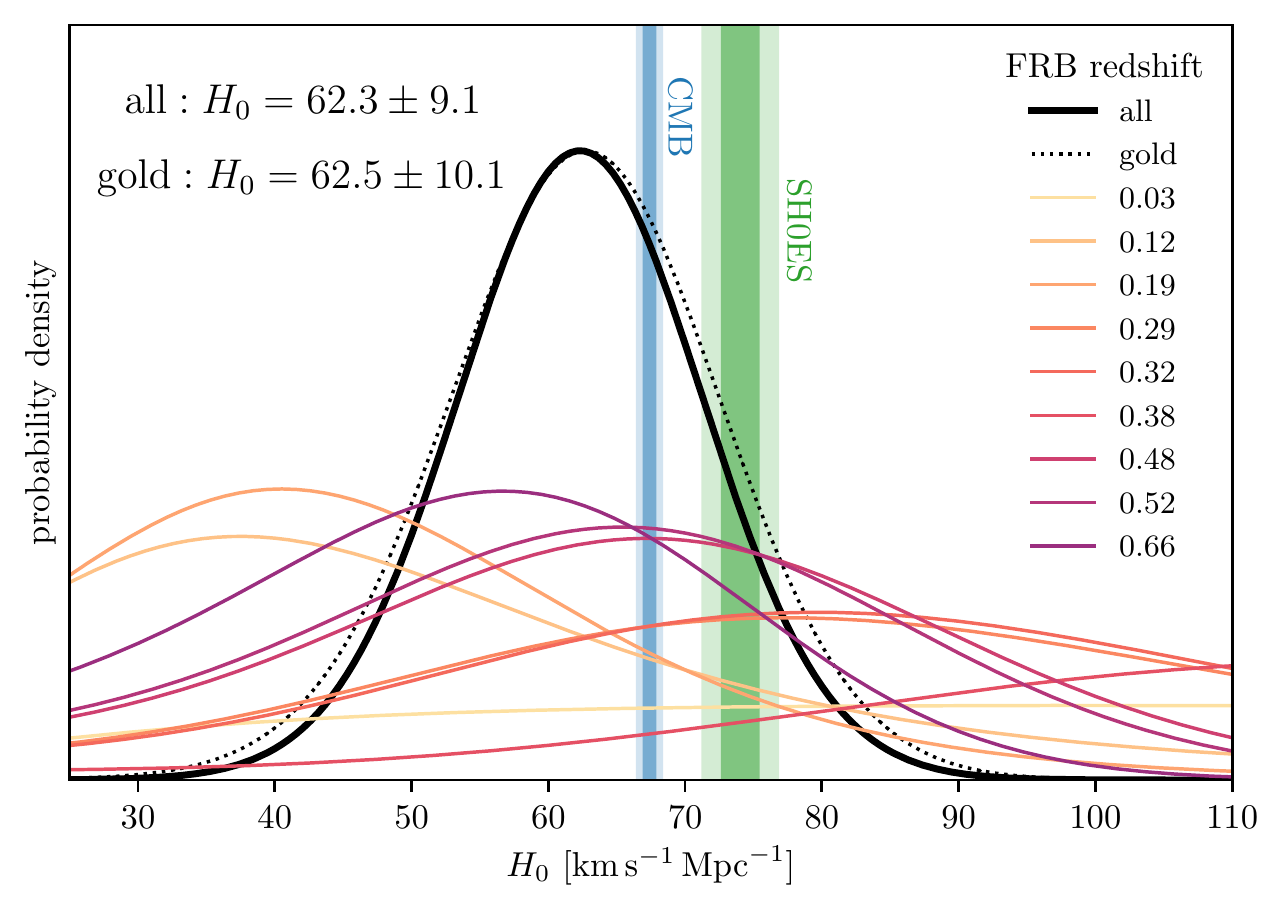}
    \caption{PDFs for the Hubble constant $H_0$ from the nine individual FRBs with known redshift (thin colored lines) and the joint constraint $H_0 = 62.3 \pm 9.1\,\rm{km}\,\rm{s}^{-1}\,\rm{Mpc}^{-1}$ of the sample (solid black). We also include the constraint from the six well-identified gold sample FRBs alone (dotted black), which results in $H_0 = 62.5 \pm 10.1\,\rm{km}\,\rm{s}^{-1}\,\rm{Mpc}^{-1}$. The $H_0$ values with $2\sigma$ error bars from Planck CMB measurements \citep[blue]{Aghanim:2018eyx} and from Cepheid-calibrated supernovae by the SH0ES collaboration \citep[green]{Riess:2019cxk} are shown as shaded bands for reference.}
    \label{fig:H0_constraint}
\end{figure}

The resulting PDFs for $H_0$ from all known events are presented in \cref{fig:H0_constraint}. We find the joint constraint from all FRBs is $H_0 = 62.3 \pm 9.1 \,\rm{km} \,\rm{s}^{-1}\,\rm{Mpc}^{-1}$, while the gold sub-sample alone yields the only marginally weaker result $62.5 \pm 10.1 \,\rm{km} \,\rm{s}^{-1}\,\rm{Mpc}^{-1}$. As can be seen from the individual constraints shown in \cref{fig:H0_constraint}, the high-redshift FRBs are by far the most sensitive data points, as the cosmological contribution to the DM becomes more important for high $z$. Low-$z$ events, on the other hand, are dominated by the DM from host and Milky Way. Especially due to the large host DM uncertainty, they have little constraining power on cosmological quantities. Since all FRBs at higher redshifts are part of the gold sample, the difference between the two results is very small.

\begin{figure}
	\includegraphics[width=\columnwidth]{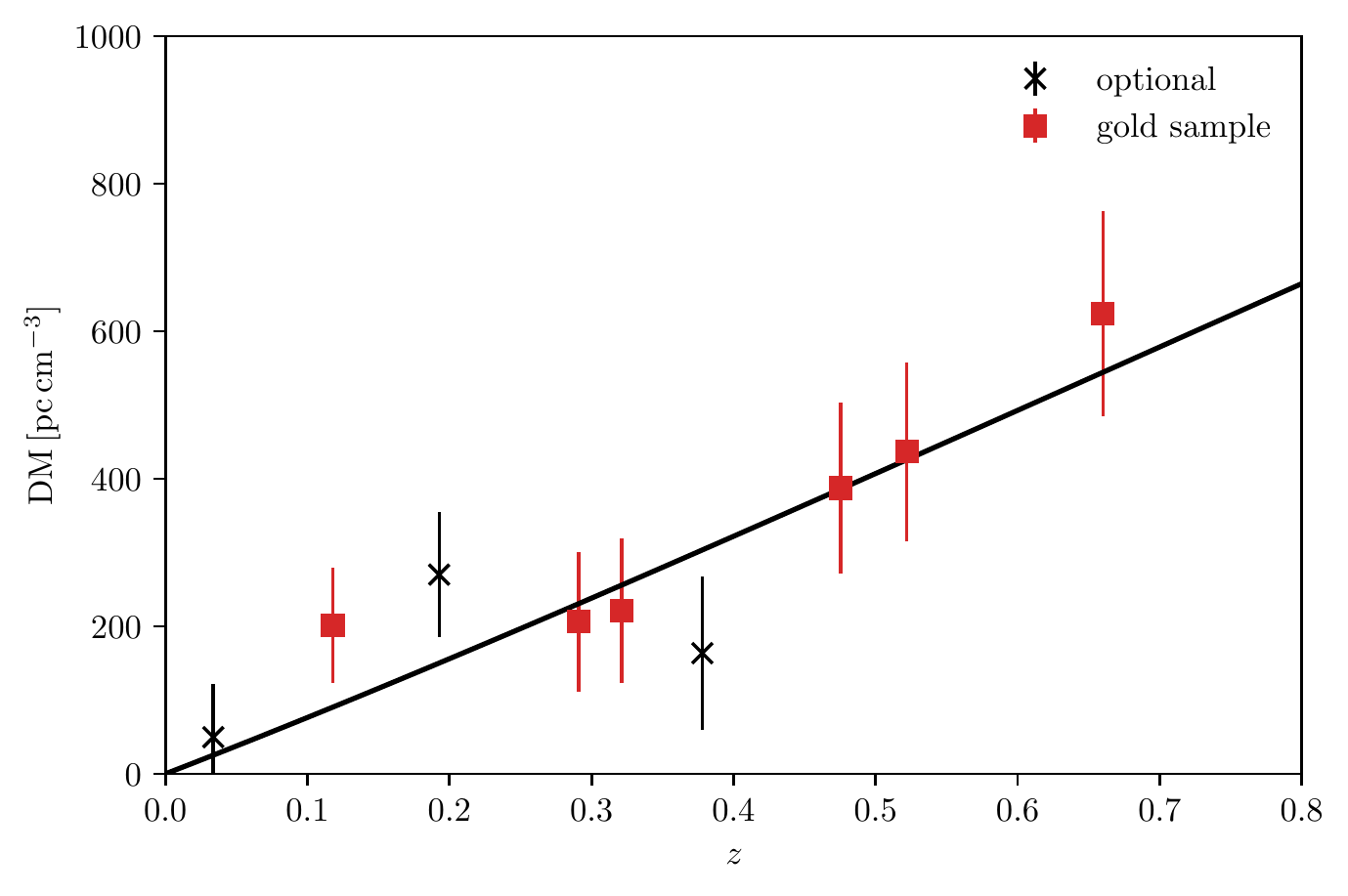}
    \caption{Best-fit DM -- redshift relation from \cref{eq:DM_LSS_v2} compared to the dispersion measure of the nine FRBs. Data points are corrected for the Milky Way contribution and the mean host DM as described in the text. The error bars include estimated scatter from the Milky Way reconstruction, the host galaxy and from the large-scale structure, as given in \cref{eq:scatter_tot}. All FRBs that are part of the gold sample, as described in the text, are marked with red squares.}
    \label{fig:DM_z_datapoints}
\end{figure}

We show the best-fit DM -- redshift relation from the joint sample in \cref{fig:DM_z_datapoints}. The error bars are based on \cref{eq:scatter_tot} since the measurement error on the individual DMs is negligible. In the plot we show the gold sample FRBs with red squares, but there is little difference in the best-fit relation when fitting only to this subset instead of the complete sample.


\section{Future Prospects}
\label{sec:forecast}

\begin{figure*}
    \centering
    \includegraphics[width=.6\textwidth]{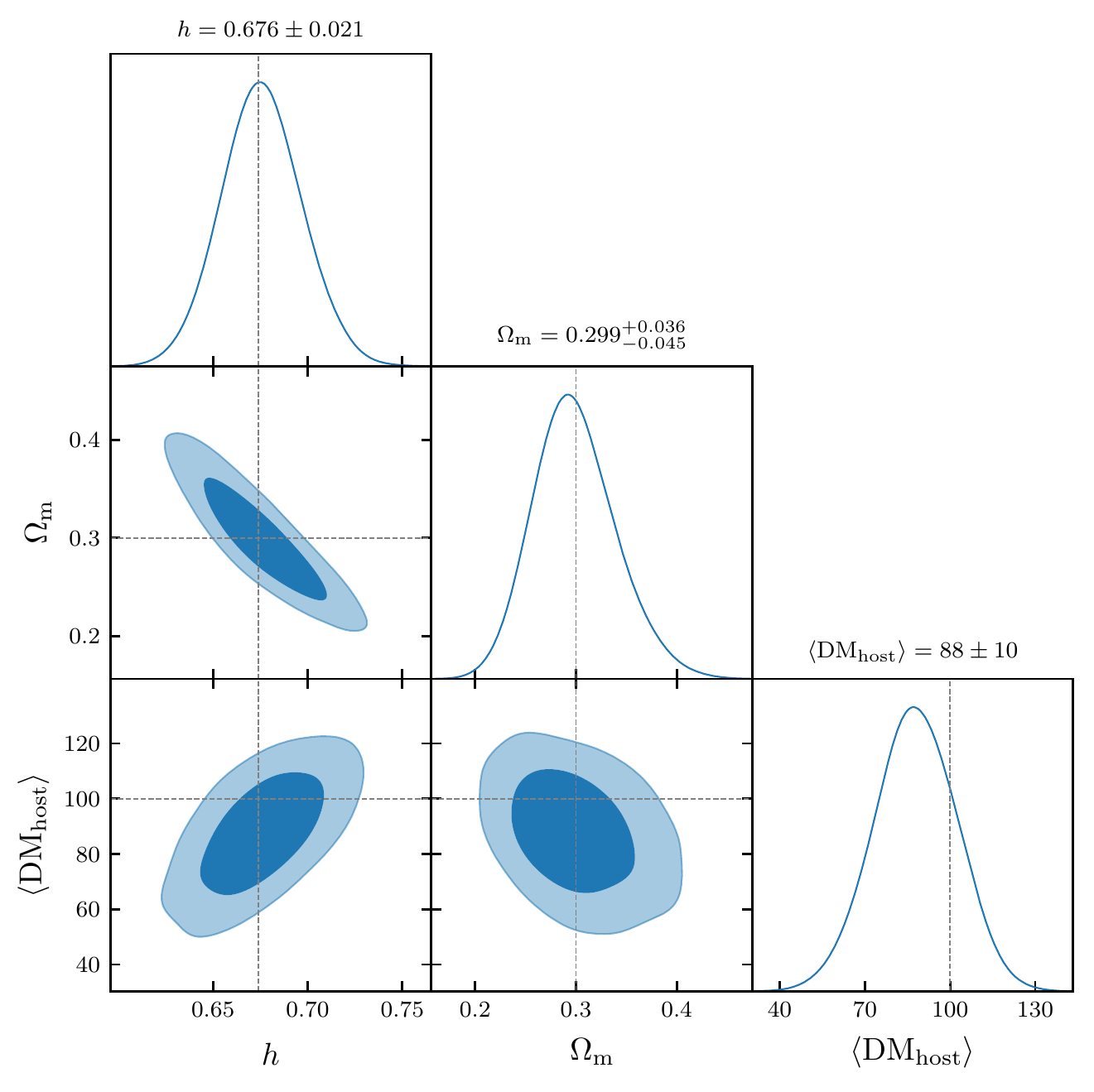}
    \caption{Expected joint constraints on the Hubble constant $h$, the matter density $\Omega_\mathrm{m}$ and the mean host DM contribution from a mock sample of 500 FRBs as described in the text. Since we are using a prior on $\Omega_\mathrm{b} h^2$, the DM -- redshift diagram is sensitive to $h^{-1}$. If we use a prior on $\Omega_\mathrm{b}$ directly, the degeneracy directions between $h$ and the other parameters are reversed. The fiducial parameter values are marked with dotted lines.}
    \label{fig:forecast_triangle}
\end{figure*}

While the limited current sample is not able to set precision constraints yet, the amount of available FRBs is expected to grow quickly over the next years. In this section, we want to determine how many events are needed for a $H_0$ measurement with per cent accuracy, which might be able to distinguish between the values preferred by early and late time measurements of $H_0$. We generate mock data from an FRB redshift distribution following the galaxy distribution of the form
\begin{equation}
\label{eq:n_z_FRB}
n(z) = z^2 \exp(-\alpha z) \: ,
\end{equation}
where $\alpha$ sets the effective depth of the sample. Given the DM of all FRB detections so far, current estimates indicate that the majority lies most likely at lower redshifts $z<1$, even though some large dispersion measures indicate detectable FRBs up to $z \sim 1.5$ or higher. To be conservative and account for the difficulty of host identification at high redshifts, we assume a very sharp cutoff with $\alpha=7$, which leads to a majority of events at intermediate redshifts $z \sim 0.3 - 0.5$. We then use inverse sampling to draw a number of events $N$ from the redshift distribution in \cref{eq:n_z_FRB}, and apply the scatter from host halo and Milky Way contributions as detailed in \cref{subsec:DM_structures}.

Given the Gaussian likelihood in \cref{eq:likelihood_single}, the error on the amplitude of the DM -- redshift relation shrinks with $N^{-1/2}$, where $N$ is the total number of data points. Extrapolating from the current uncertainty of the measurement in \cref{sec:results}, the one-dimensional $H_0$ constraint reaches per cent accuracy with a few hundred of data points. The main shortcoming of the current measurement lies in the assumptions about the mean host DM, which shifts the overall observed DM and is therefore difficult to distinguish from $H_0$ with a small sample. However, since the mean host halo DM scales with $(1+z)^{-1}$, more data can reliably tell the effects apart and determine both parameters separately.
For the forecast, we therefore add the mean host contribution $\langle \mathrm{DM}_\mathrm{host}\rangle$ as a free parameter determined by the data.
A large sample could also allow binning of the FRBs according to the type of the host, since different galaxy types will likely differ in their contributions to the intrinsic DM. This can then result in smaller scatter in the host DM distribution and increase the accuracy of the final constraints, but we will not include this approach in our forecast.

To arbitrate the tension between early and late time measurements of the expansion rate, we aim for a FRB $H_0$ measurement fully independent of CMB data. Since the DM -- $z$ relation is sensitive to $\Omega_\mathrm{b} h$, the degeneracy has to be broken by an external measurement. We will assume a Gaussian prior determined from Big Bang Nucleosynthesis (BBN)\footnote{This still assumes standard physics at the time of BBN, and in particular no additional relativistic species in the early Universe.} on the baryon density, $\Omega_\mathrm{b} h^2 = (2.235 \pm 0.037) \times 10^{-2}$ \citep{Cooke:2017cwo}, and jointly fit for $H_0$, $\Omega_\mathrm{m}$, $\Omega_\mathrm{b}$ and $\langle \mathrm{DM_\mathrm{host}} \rangle$ with flat priors on th remaining parameters. We assume a total of $N = 500$ events and explore the parameter space using the likelihood function in \cref{eq:likelihood_tot} with the help of the \texttt{emcee} sampler \citep{Foreman_Mackey:2013PASP}.

We show the results in \cref{fig:forecast_triangle}. Even after marginalising, a sample of ~500 FRBs with optical counterparts can be sufficient to set competitive constraints on $H_0$. Note that since we are using a BBN prior on the physical density $\Omega_\mathrm{b} h^2$, the measurement is again sensitive to $h^{-1}$ and the intuitive degeneracy directions of parameters in \cref{fig:forecast_triangle} are reversed: the constraint on the Hubble constant and mean host DM become positively correlated, and $H_0$ and $\Omega_\mathrm{m}$ are negatively correlated, even though a large $\Omega_\mathrm{m}$ leads to shorter distances for a given redshift. If instead we use a prior on $\Omega_\mathrm{b}$ directly, for example with the help of measurements of the gas fraction in massive galaxy clusters \citep{Allen:2008MNRAS}, together with other LSS probes to determine $\Omega_\mathrm{m}$, the degeneracy direction would be reversed.

\section{Conclusion}
\label{sec:conclusion}

We present a new measurement of the Hubble constant $H_0$ based on the dispersion measure -- redshift relation of fast radio bursts (FRBs). The method is similar to the determination of $H_0$ from the luminosity -- redshift relation of calibrated SN Ia. 
The total DM is dominated by the cosmological signal for redshifts $z>0.3$. With the current small sample of nine FRBs with known host galaxies, we constrain the Hubble constant to $H_0 = 62.3 \pm 9.1 \,\rm{km} \,\rm{s}^{-1}\,\rm{Mpc}^{-1}$. We also limit the analysis to the six events with the most reliable host identification (the gold sample), with an almost identical result of $H_0 = 62.5 \pm 10.1\,\rm{km}\,\rm{s}^{-1}\,\rm{Mpc}^{-1}$, since most of the excluded FRBs are located at low redshifts. The current main limitations lie in the very small number of available events with sufficient localisation and in the uncertainty about the DM contribution from the host galaxy. Both of these can be solved by a larger sample of localised FRBs. In fact, dedicated searches with excellent angular resolution are expected to detect hundreds of bursts and their host galaxies over the next years.

We demonstrate with a forecast that, with the data available in the near future, it is possible to set precision constraints on $H_0$ fully independent from the CMB or other cosmological measurements, while simultaneously determining the stochastic host halo contribution. Since the cosmological and the stochastic contributions to the DM scale differently with redshift, a sample of a few hundred FRBs can reliably distinguish the two effects.
This demonstrates the potential of FRBs for precision measurements of cosmological parameters.

\section*{Acknowledgements}
We thank the authors of \cite{Macquart:2020lln} for making their numerical tools for FRB-related calculations publicly available. SH would like to thank Raffaella Capasso for helpful remarks about the manuscript.

 SH acknowledges support from the Vetenskapsr\r{a}det (Swedish Research Council) through contract No.\ 638-2013-8993 and the Oskar Klein Centre for Cosmoparticle Physics. RR is supported by the European Research Council (Grant No. 770935). RL acknowledges support by a Technion fellowship.

\section*{Data Availability}

The FRB data for our analysis is available publicly at \href{http://frbcat.org/}{http://frbcat.org/} \citep{Petroff:2016tcr}. The mock data for our forecasts can be generated as described in the main text.



\bibliographystyle{mnras}
\bibliography{bibliography}



\bsp	
\label{lastpage}
\end{document}